\documentclass[final]{aipproc}
\layoutstyle{6x9}

\begin{document}

\title{Triatomic Molecular Systems and Three-body forces: The Ar$_3$ case}
\author{M. L. Lekala}
{address={
Physics Department, University of South Africa, PO Box 392, 
    Pretoria 0001, South Africa}}
\author{S. A. Sofianos}
{address={
Physics Department, University of South Africa, PO Box 392, 
    Pretoria 0001, South Africa}}

\begin{abstract}
We performed bound state calculations to obtain the first few vibrational 
states for the Ar$_3$ molecular system. The equations used are of Faddeev-type
and are solved directly as three-dimensional equations in 
configuration space, 
i.e. without resorting to an explicit partial wave decomposition. 
In addition to realistic pairwise interactions, we employ long range 
three-body forces. Our  results are in good agreement 
with those obtained by other methods based on partial wave 
expansion  and show a significant contribution of the three-body forces 
($>$10\%) to the binding energy and thus their inclusion is, 
in general, warranted in studying similar triatomic systems. 
\end{abstract}

\maketitle

\section{Introduction}
%%%%%%%%%%%%%%%%%%%%%%

In calculations of the vibrational spectra of triatomic inert gases, 
pairwise interactions are used as a first approximation. 
Three-body interactions are usually neglected on the assumption that their 
contribution is small, or on the ground that they are poorly known. For 
example, 
in the case of $^4$He trimer the contribution of the three-body forces to 
the binding is known to be small -- of the order of $< 1$\% \cite{Esry}.
In the case of Ar$_3$ system, however, it has been shown that the inclusion of 
three-body forces substantially improves the agreement between 
the third virial coefficient and  experiment~\cite{johnson}. In view of the 
fact that realistic two-body forces and rigorous theoretical approaches are 
nowadays available, a systematic investigation of the role of three-body 
forces in trimers is warranted. \\

In the present work we are concerned with the calculation of the vibrational 
spectrum of the Ar$_3$ with and without three-body forces. The choice of this 
system is ideal for the following 
reasons: First, realistic inter-atomic forces are available~\cite{aziz}
 and therefore the role of three-body forces can be 
investigated unambiguously; second, unlike  nuclear three-body systems,
Ar$_3$ is a bosonic one and thus complications due to spin
and isospin are absent. Therefore, effects form 
numerical inaccuracies can be minimized. Furthermore, the  Ar$_3$ is 
known to exhibit the so-called ``liquid-like'' behavior making it a good 
candidate for investigating other phenomena such as recombination processes 
at ultra-low energies, low-temperature crystallization, 
photochemistry,  etc.  Moreover, the behavior of this trimer under 
certain external thermodynamical conditions especially those close to phase 
transitions are of extreme importance in understanding the influence of 
three-body forces on the gas-liquid coexistence of Argon. \\

In our investigations we employ three-body Faddeev-type  equations 
in configuration space obtained within the framework 
of the total-angular-momentum representation~\cite{sofi}. 
These  equations are solved as three-dimensional equations, 
{\em i.e.} without resorting to explicit partial 
wave decomposition and thus the question on whether enough partial 
waves are included or not is avoided.

\section{Our Approach}
%%%%%%%%%%%%%%%%%%%%%%

In the presence of both two- and three-body forces, the three-particle 
Hamiltonian reads~\cite{sofi}
\begin{equation}
      {\rm H}_{\rm 3B} = {\rm H}_0 + \sum_{i=1}^3V_i^{\rm (2B)} + V^{\rm (3B)}
\label{hamiltonian}
\end{equation}
leading to a set of coupled differential Faddeev-type equations
($\hbar^2/2\mu=1$)
\begin{equation}
\biggl[   {\rm H}_0 + 
      V_i^{\rm (2B)}({\bf x}_i) + V^{\rm (3B)}({\bf x}_i,{\bf y}_i) 
    - E_{\rm 3B}
\biggr]     \Phi_i({\bf x}_i,{\bf y}_i) = 
    - V_i^{\rm (2B)}({\bf x}_i)\sum_{j\not= i}
      \Phi_j({\bf x}_j,{\bf y}_j)\,,
\label{faddeev}
\end{equation}
where ${\rm H}_0$ is the free Hamiltonian, $V_i^{\rm (2B)}$
 and $V^{\rm (3B)}$  are the two-  and three-body potentials
respectively,  
$E_{\rm 3B}$ the binding energy, $({\bf x}_i\,,{\bf y}_i)$ 
the Jacobi coordinates, and $\Phi_i\,\,(i=1,2,3)$ the Faddeev 
components. 
In the case of zero total angular momentum, and for identical particles, 
H$_0$ is given by
\begin{equation}
{\rm H}_0  =  - \frac{\partial^2}{\partial {x}^2} - 
          \frac{\partial^2}{\partial {y}^2} - 
          \left(\frac{1}{x^2} + \frac{1}{y^2}\right)
          \frac{\partial}{\partial {z}}(1 - {z}^2)
          \frac{\partial}{\partial {z}}
\label{hzero}
\end{equation}
where $x = |{\bf x}|$ is the usual two-body Jacobi coordinate,
 $y = |{\bf y}|$ the 2+1 coordinate, and $z = \cos({\bf x}\cdot{\bf y})$.
In order to obtain a solution, the energy $E$ is treated as a 
parameter and thus Eq. (\ref{faddeev}) is transformed into an 
eigenvalue equation
\begin{equation}
      (E{\rm I} - H_0 - {\cal{V}})^{-1}P\Phi = \Lambda\Phi
\label{eigenvalue}
\end{equation}
where $P$ is the total permutation operator of the coordinate variables, and 
$\cal{V}$ contains all potential terms (details on the formalism can be found in Ref. 
\cite{sofi}).  We solve the three-dimensional 
Eq.~(\ref{eigenvalue}) iteratively using the Arnoldi-type method.  
Physical solutions correspond to the cases when the 
eigenvalue $\Lambda =1 $. Thus the problem of calculating the vibrational 
binding energies is reduced to finding the discrete spectrum of the  
operator 
\begin{equation}
{\cal{L}} = (E{\rm I} - H_0 - {\cal{V}})^{-1}P\,.
\label{faddeevop}
\end{equation}

\section{Results and Discussion}
%%%%%%%%%%%%%%%%%%%%%%%%%%%%%%%%%

We calculated the binding energies for the ground state and first 
excited  state of  the  Ar$_3$ trimer by employing  two variant 
highly repulsive Ar-Ar realistic
 potentials of Aziz~\cite{aziz}. For the three-body force we use 
the triple-dipole Axilrod-Teller type~\cite{axilrod} interaction. 
The results obtained using  pairwise 
forces (2BF) and a combination of two- and 
three-body forces (2BF + 3BF) are shown in  Table~\ref{results}. It 
is seen that the results with both two- and three-body interactions 
are in fair agreement with, for example, those of Refs.~\cite{leitner} and 
\cite{cooper}, via different formalisms. It is clear
that the  Faddeev-type formalism employed is
suitable in studies concerning three-molecular systems and 
the question of handling the multitude of
couple equations for the various partial waves
can be avoided by solving the  three-dimensional
equations directly. Such an approach is highly desirable
when strong repulsive two-body forces are involved. The practically 
hard repulsive core nature of the inter-molecular van der Waals forces gives
rise to strong  correlations which require a large number of partial
waves to obtain converged results. Finally, we mention 
that our results show a significant contribution to the binding 
from three-body forces ($>$10\%). In this regard the inclusion of three-body 
calculations in triatomic molecular systems is, in general, warranted. 
Further compilation of results are underway.

\vspace*{1cm}

\begin{table}[h]
\setlength{\AIPhlinesep}{0pt}
\centering
\begin{tabular}{|l|c|c|c|c|}                                         \hline
\tablehead{1}{c}{c}{Potential}&
\tablehead{2}{c}{c}{This work}&
\tablehead{2}{c}{c}{Other works}                                   \\\hline
&\hspace*{0.5cm} 2BF\hspace*{0.5cm}
&\hspace*{0.5cm}2BF + 3BF\hspace*{0.5cm}
&\hspace*{0.5cm}Ref. \cite{leitner}\hspace*{0.5cm} 
&\hspace*{0.5cm}Ref. \cite{cooper}\hspace*{0.5cm}                  \\\hline
   \hspace*{0.5cm}HFD-B2\hspace*{0.5cm}&\hspace*{0.5cm}-0.0329   
&  \hspace*{0.5cm}-0.0356              &\hspace*{0.5cm} -0.0314   &\\
&  \hspace*{0.5cm}-0.0297              &\hspace*{0.5cm}-0.0301    & 
   \hspace*{0.5cm}-0.0286              &          \\\hline
   \hspace*{0.5cm}HFD-C\hspace*{0.5cm} &\hspace*{0.5cm}-0.0327    &    
   \hspace*{0.5cm}-0.0363              &        & \hspace*{0.5cm}-0.0316  \\
&  \hspace*{0.5cm}-0.0295              &\hspace*{0.5cm}-0.0325    &
&  \hspace*{0.5cm}-0.0278                                          \\\hline
\end{tabular}
\caption{The ground and first excited  states for the 
Ar$_3$ obtained with  2BF and 2BF+3BF forces. 
The results of Ref. \cite{leitner} and \cite{cooper} are with 2BF+3BF
forces. Energies are given in eV.}\vspace*{0.1cm} 
\label{results}
\end{table}

\begin{theacknowledgments}
One of us, M.L.L., would like to acknowledge the financial support by  
the National Research Foundation under Grant number: NRF GUN 2054317.
\end{theacknowledgments}

\end{document}